# Exotic Nd 4*f* electron magnetism in Nd$_2$RhSi$_3$

E V Sampathkumaran[1, *, #], Kartik K Iyer[2], Sudhindra Rayaprol[3] and Kalobaran Maiti[2]

[1]Homi Bhabha Centre for Science Education, Tata Institute of Fundamental Research, V. N. Purav Marg, Mankhurd, Mumbai – 400088, India

[2]Tata Institute of Fundamental Research, Homi Bhabha Road, Colaba, Mumbai – 400005, India

[3]UGC-DAE Consortium for Scientific Research - Mumbai Centre, BARC Campus, Trombay, Mumbai – 400085, India

[*]Corresponding author email: sampathev@gmail.com

**Abstract**

The compound, Nd$_2$PdSi$_3$ belonging to $R_2$PdSi$_3$ family ($R$ = rare-earths) has been known to exhibit exotic behavior due to unusual Nd 4*f*-hybridization. Here, we study the electronic properties of Nd$_2$RhSi$_3$ employing ac and dc magnetization, heat-capacity, electrical resistivity and magnetoresistance measurements, as Nd 4*f*-hybridization is expected to be slightly different due to the changes in the 4*d* element. The experimental results establish that, like the Pd analogue, this compound also exhibits ferromagnetic ordering at a rather high temperature of 16.5 K, unlike many other rare-earth members of this family which are antiferromagnetic, with a complex magnetism at further lower temperatures (< ~10 K). There are differences in the measured properties with respect to the Pd analogue, the most important one being the observation of spin-glass features at a temperature significantly higher than the Curie temperature. This is attributed to a gradual evolution of cluster magnetism with decreasing temperature. We infer that the changes in Nd 4*f* hybridization due to Rh 4*d* (instead of Pd 4*d*) plays a role in some fashion for such differences. These properties are attributed to the competing magnetic ground states due to geometrical frustration arising out of triangular arrangement of Nd ions with a bond disorder.

## 1. Introduction

It is well known that the radial extension of 4*f* orbitals reduces with the increase in atomic number. Therefore, the degree of 4*f* extension, and consequent 4*f* hybridization gradually reduces as one traverses from Ce to Sm. While the role played by such 4*f* hybridization on magnetic ordering behavior is well recognized for Ce intermetallics in the literature, corresponding influence of partial 4*f* radial extension on magnetic ordering has been less demonstrated for other light rare-earth systems. In this respect, the ternary rare-earth compounds of the type, $R_2$PdSi$_3$, derived from the ordered replacement of boron site in the (hexagonal) AlB$_2$ by Pd and Si [1] have been of great interest. This family has been investigated by several macroscopic and microscopic experimental methods for the past three decades [2-23] and was found to show an unusual Nd 4*f* hybridization effect on magnetism [17], apart from various other exotic magnetic phenomena like Kondo-lattice anomalies (in the Ce compound), large magnetoresistance (MR) even in the paramagnetic state of heavy *R* members, magnetocaloric effect, spin-glass freezing (due to layered triangular network of *R* layers) including reentrant behavior, and signatures of topological Hall effect in Gd$_2$PdSi$_3$ (as early as 1999, see Ref. 5), and magnetic skyrmion behavior [22-24]. Superconductivity was also reported in the isostructural $Y_2PdGe_3$ [25] and $Y_2PtGe_3$ [26]. In $Nd_2PdGe_3$, there is an enhancement of magnetic ordering temperature [25] with respect to de Gennes scaled value, which is an indication of anomalous magnetism. Exhaustive studies have been performed recently on $Nd_2PdSi_3$, exhibiting exceptional magnetic behavior within the $R_2PdSi_3$ due to the onset of ferromagnetism (at about 16 K which is rather high, disobeying Gennes scaling), whereas other members of $R_2PdSi_3$ order antiferromagnetically. Such an exotic magnetism, uncommon for Nd compounds, has been found [17] to be due to subtlety in the hybridization of Nd 4$f$ states with the valence electrons in the PdSi-layer [18]. Additionally, the magnetism of this Nd-Pd compound also has been revealed by neutron diffraction and muon spin rotation studies [19] to be exotic due to a competition of a sinusoidally modulated antiferromagnetic component with ferromagnetism as a function of temperature ($T$).

The radial extension of Rh 4$d$-orbitals is expected to be slightly larger than Pd 4$d$-orbitals, which provides a good platform for the study of the role of differences in the Nd 4$f$-hybridization on the electronic properties in comparison with the Pd-analogue. Rh-based materials, namely, $R_2RhSi_3$, were discovered a few decades ago [27, 28], but were not paid similar attention subsequently. There were some efforts in recent years for $R$ = Ce, Gd, Tb, Dy, Ho and Er, which were shown to exhibit different kinds of magnetic anomalies due to geometrical frustration [29-39]. The focus in this article being $R$ = Nd, it may be stated that very little work has been done on $Nd_2RhSi_3$ as well. Initial $dc$ magnetization ($M$) report in 1984 [27] indicated that $Nd_2RhSi_3$ behaves differently with respect to other members in the Rh family also, suggestive of ferromagnetic order around 15 K, as in the Pd series.

A neutron diffraction study at 4.2 K suggested spiral ferromagnetic structure for this compound at this temperature [40]. In this article, we present the results of detailed investigations on the magnetic properties of this compound.

The ternary family of $R_2RhSi_3$ is made up of Rh-Si and $R$ layers stacked along $c$-direction, as shown in Figure 1a. Rh-Si forms a honeycomb network, intercalated by $R$ layers with $R$ forming triangles. As brought out in Ref. 27, there are multiple bond distances within a layer for a given pair of elements, as shown in Figure 1b for the Nd network. The figure 1 is drawn using VESTA program [41]. There is a doubling of unit-cell parameters as indicated by superstructure lines. Till now it has not been possible to resolve the space group whether it is $P\bar{6}2c$ or $P6_3/mmc$. Our detailed investigations by $ac$ and $dc$ magnetization, heat-capacity ($C$), electrical resistivity ($\rho$) and magnetoresistance measurements reported here show interesting behavior which is very different from other materials of this class.

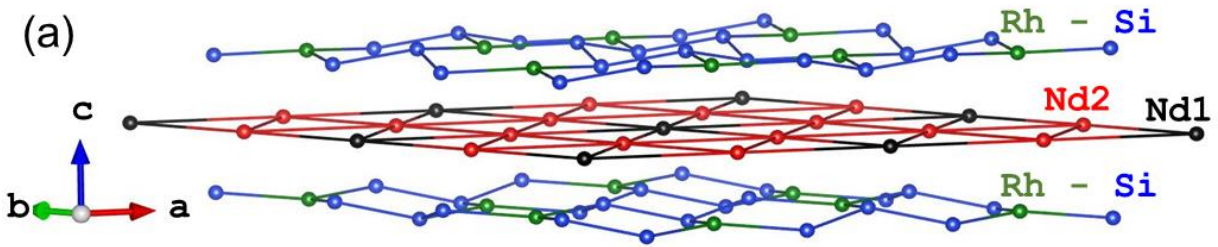

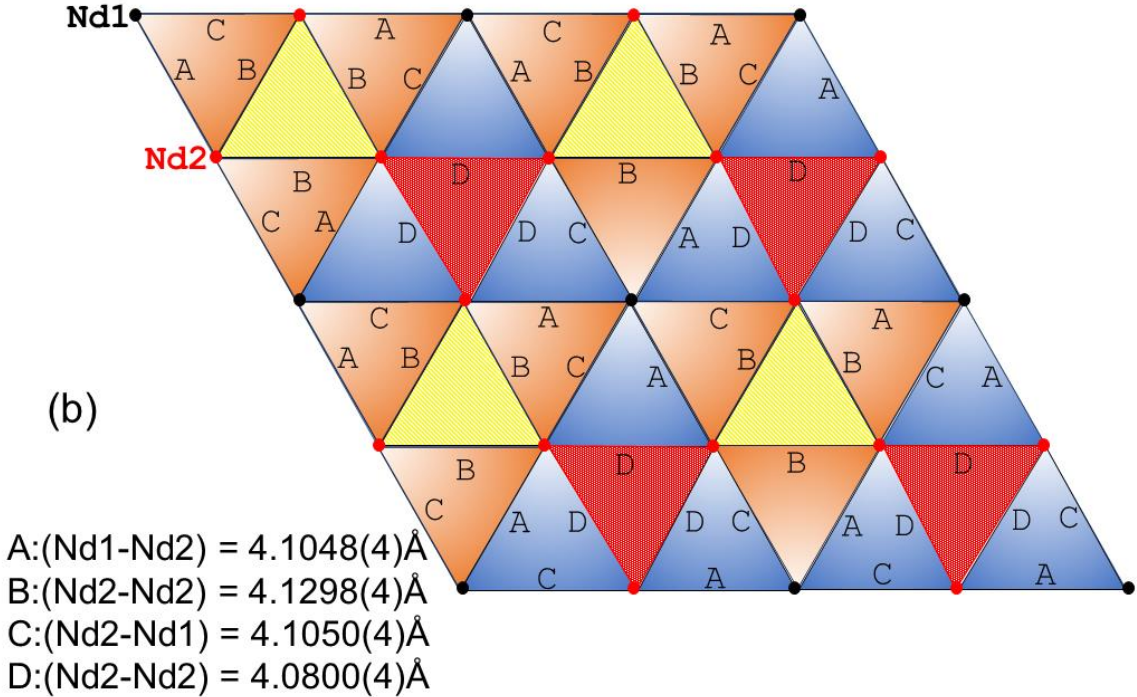

**Figure 1**. (a) A view of Rh-Si layers and rare-earth layers of Nd$_2$RhSi$_3$, stacked along *c*-axis, for instance, for *P*6$_3$/*mmc* space group. (b) A view of the triangular network of *R* ions in one layer along *c*-axis is shown. The bond distances are also mentioned. Triangles of the same dimensions are shaded with the same color.

## 2. Experimental Details

High quality sample was prepared in the polycrystalline form by arc melting stoichiometric amounts of high purity constituent elements, Nd (>99.9%), Rh (>99.99%) and Si (>99.999%), in an argon atmosphere. The molten ingot was annealed at 850°C in an evacuated sealed quartz tube for 1 week. The sample thus prepared was ascertained to be in single phase by x-ray (Cu-K$\alpha$) diffraction. The x-ray diffraction pattern with Rietveld analysis is shown in figure 2 for both the space groups, with fitting performed using the position coordinates mentioned in Ref. 27 as the starting model. The appearance of superstructure lines confirms doubling of the unit-cell with respect to the (parent) AlB$_2$ structure. The lattice constants obtained by Rietveld fitting are in good agreement with Ref. 27. A careful look at the simulated pattern for the space group *P*$\bar{6}$2*c* (bottom panel of Fig. 2) shows that some Bragg peaks (in 2θ) say at 16.4º, 35º and 39.5º, exhibit additional intensities which are not present in the raw data. These Bragg positions are allowed by *P*6$_3$/*mmc* space group too, but their intensity is almost zero in simulated pattern and hence matches well with observed XRD data. Therefore, we tend to believe that the *P*6$_3$/*mmc* space group may be the more appropriate one.

Other bulk measurements, specified below, were performed utilizing commercial instruments, procured from Quantum Design. With the help of superconducting quantum interference device, the *T*-dependence of magnetic susceptibility (χ) was obtained in a field of 5 kOe for zero-field-cooled (ZFC) condition of the specimen in the range 1.8-300 K; in addition, the measurements were performed for ZFC as well as field-cooled (FC) conditions below 60 K with a field of 100 Oe. Isothermal *M* curves were obtained at several temperatures in the low *T* range using a vibrating sample magnetometer. Isothermal remnant magnetization ($M_{IRM}$) behavior at selected temperatures was studied by a standard protocol - that is, after cooling the specimen in zero-field to the desired temperature, the specimen was left in a field of 5 kOe for 5 minutes; $M_{IRM}$ curves as a function of time (*t*) were obtained after switching of the field. Heat-capacity behavior in the range 1.8-130 K was obtained in 0, 10, 30, 50 and 100 kOe using a Physical Properties Measurements System; ac $\chi$ measurements (< 100 K) with 4 frequencies ($v$ = 1.3, 13, 133 and 1339 Hz) and dc electrical resistivity measurements (<300 K) in zero-field as well as in-field were performed with the same instrument.

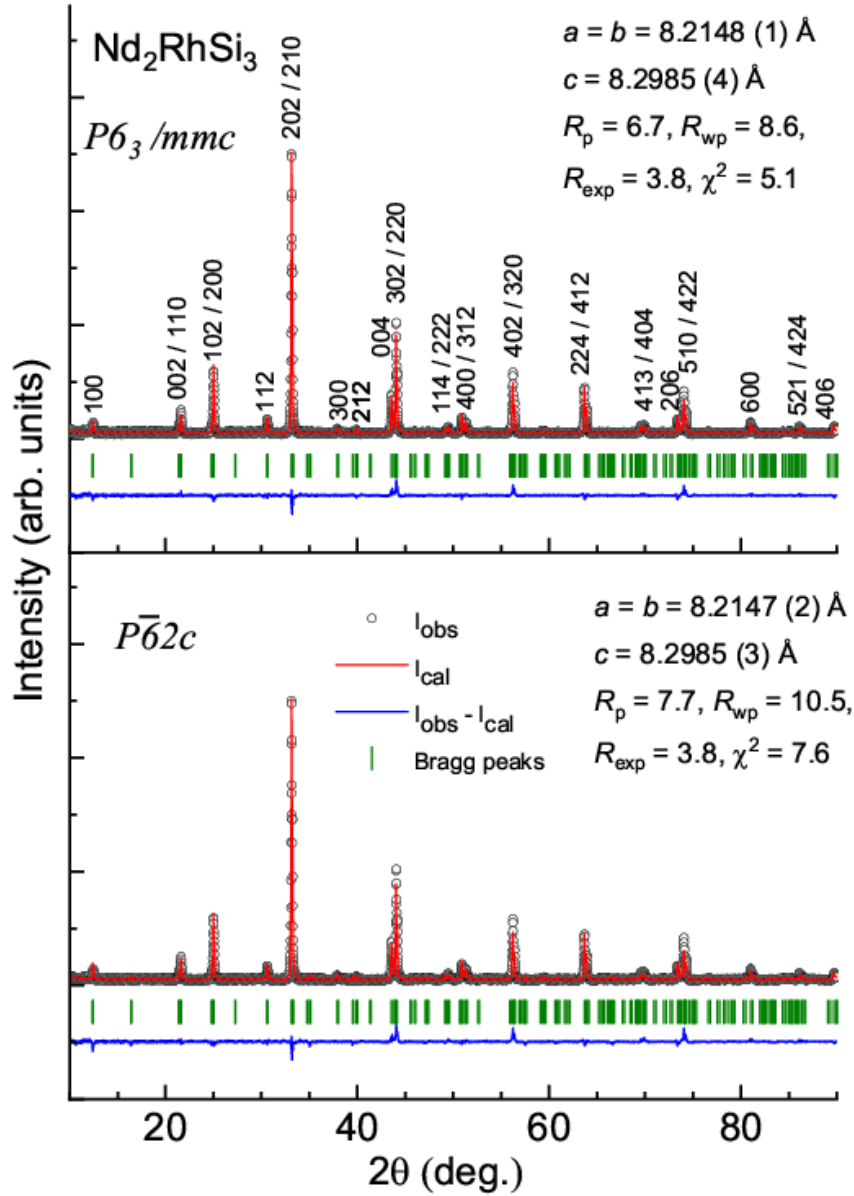

**Figure 2**. Powder x-ray diffraction pattern of $Nd_2RhSi_3$ obtained using Cu K$\alpha$ radiation. Rietveld fitted results are shown for both the space groups. *R*-factors and lattice constants are shown.

## 3. Results

*3.1 dc magnetization*

The results of dc magnetization measurements in a field of 5 kOe are shown in figure 3a. The plot of inverse $\chi$ versus $T$ is linear above 50 K, and there is a gradual deviation from the linearity due to the crystal-field effects. The gradual formation of magnetic clusters (*vide infra*) would also add to this. The value of the effective moment ($\mu_{eff}$) obtained from the linear region is ~3.7 $\mu_B$/Nd, which is very close to that for free trivalent Nd ion (3.6 $\mu_B$). The value of the paramagnetic Curie temperature ($\theta_p$) is interestingly close to zero (0.7 K), which implies that either magnetic exchange interaction strength is very weak or the ferromagnetic and antiferromagnetic components coexist with similar magnitudes of strength. The low temperature plot (<50 K), shown in the left inset of figure 3a, reveals that there is a sharp upturn in $\chi$ below ($T_C$ =) 16.5 K, as brought out by the derivative curve (right inset). This is due to the onset of long-range magnetic order.

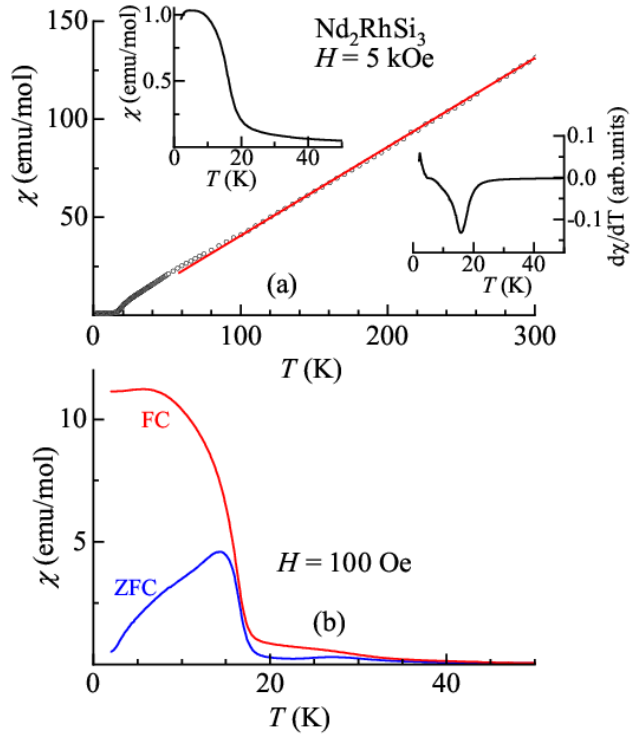

**Figure 3**. (a) Inverse of magnetic susceptibility ($\chi$) of Nd$_2$RhSi$_3$, obtained in a field of 5 kOe, and a line shows Curie-Weiss fitting above 50 K. Inset shows the $\chi$ (for zero-field-cooled condition) and its derivative below 50 K. (b) $\chi(T)$ curves obtained in 100 Oe for zero-field-cooled and field-cooled conditions below 50 K.

Such a large value of magnetic ordering temperature with respect to that of $\theta_p$ suggests that the former possibility is not a valid one. In order to gain further knowledge of the magnetically ordered state, we have measured magnetization in a low-field of 100 Oe for ZFC and FC conditions by cooling the sample from 50 K, the results of which are shown in figure 3b. In conformity with the data taken in 5 kOe, there is an upturn below 16.5 K, both in ZFC and FC curves. While ZFC curve peaks at 14.5 K, the FC curve separates, a characteristic feature of spin-glasses [42]. However, FC curve, rather than remaining flat (as in conventional spin-glasses) with decreasing *T* below 14.5K, keeps increasing, as in Pd analogue, before attaining flatness. Such an upturn in FC curve is typical of cluster spin-glasses [see, for instance, Refs. 43-48].

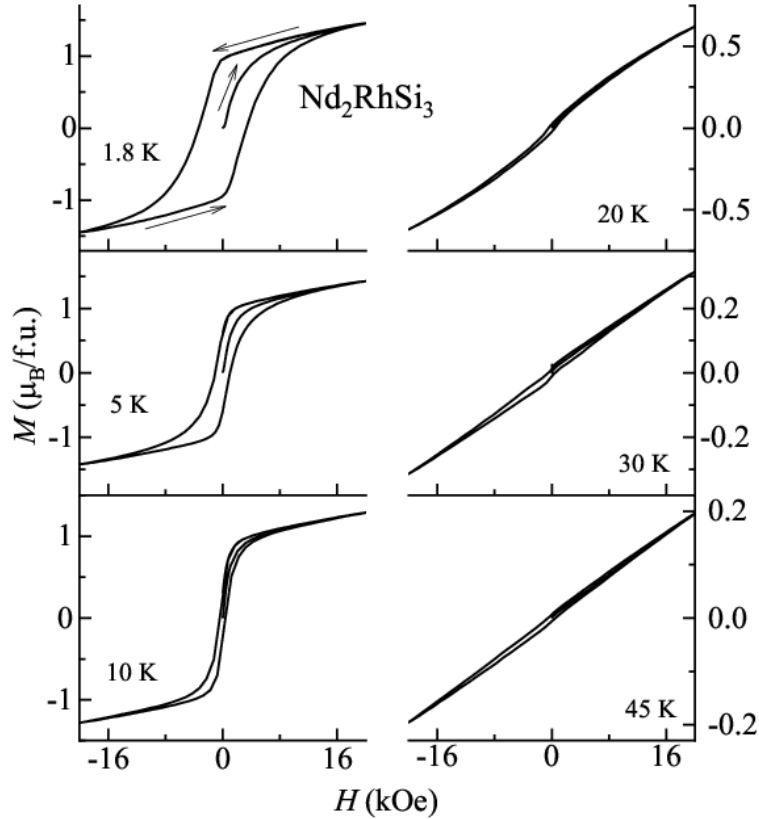

**Figure 4**. Magnetic hysteresis loop behavior of $Nd_2RhSi_3$ in the range (-20 to 20 kOe) at 1.8, 5, 10, 20, 30 K, and 45 K. Arrows for 1.8 K are shown to track the path of the data when the magnetic field is increased.

There are additional noteworthy features in the $\chi$ data, which are absent in $Nd_2PdSi_3$: Careful measurements under ZFC condition reveal that there is a weak shoulder at about 10 K, as though there is an additional magnetic feature. This is also sensitive to the strength of the magnetic field, as it is shifted to a slightly lower temperature when measured with 5 kOe. See the inset of figure 3a. ZFC-FC curves actually do not merge up to about 40 K, as though there are strange magnetic correlations persisting far above $T_C$, as shown in Figure 3b.

We have measured isothermal magnetization behavior up to 70 kOe (measured for ZFC condition) and the results are shown in figure 4 in the form of hysteresis loops (seen below 20 kOe only) at various temperatures. There is no worthwhile feature above 20 kOe, except that the curves do not saturate up to the highest measured field, even at 1.8 K; the highest value of the magnetic moment (~1.5$\mu_B$ per formula unit near 70 kOe), is far below that expected for trivalent Nd ion. These findings imply that there is an antiferromagnetic component in zero field and/or magnetic-field induced occupation of higher lying crystal field levels. While these observations are similar to that in $Nd_2PdSi_3$, the behavior of the hysteresis loop is different from the Pd case, in the sense that the virgin curve lies well inside the loop in the low temperature range, as though the ferromagnetic correlations dominate over antiferromagnetic correlations. Note that the *M*(*H*) curves are weakly hysteretic even above $T_C$, say at 30 K, which suggests that the ferromagnetic clusters form before long range magnetic ordering sets in. The hysteresis gradually disappears with increasing temperature (see, the curve for 45 K), with a simultaneous decrease of coercive field; the plot was found to be linear at 50 K.

*3.2 Heat-capacity*

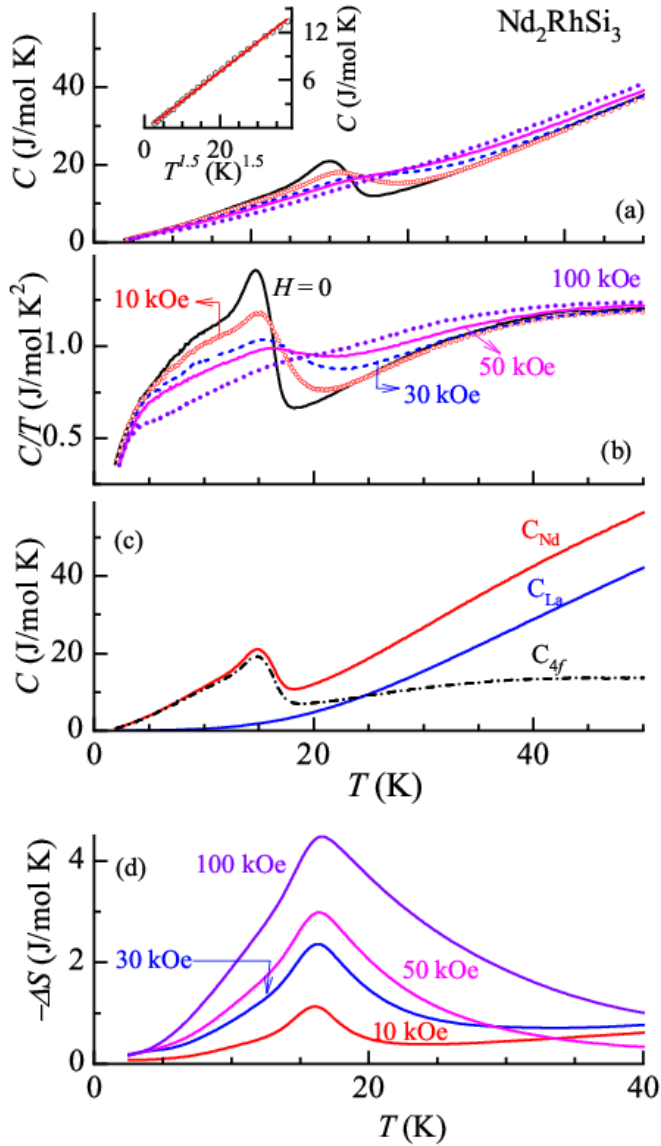

**Figure 5**. (a) Heat-capacity (*C*) as a function of temperature (<40 K) for $Nd_2RhSi_3$, measured in the presence of external magnetic fields as well. A $T^{1.5}$ form at low temperatures is also shown with the continuous representing the fit. (b) Heat-capacity divided by temperature curves. (c) Zero-field heat-capacity as a function of temperature for $Nd_2RhSi_3$ and $La_2RhSi_3$ and the 4*f* contribution to heat-capacity for the former. d) Isothermal entropy changes while varying the magnetic field from zero to a finite value, as shown in the figure below 40 K.

In order to throw more light on the nature of magnetic ordering, we show the heat-capacity behavior as a function of *T* below 50 K in figure 5. In the zero-field curve (Figure 5a), following a gradual decrease with *T*, there is an upturn at 18 K, followed by a peak at 15 K, with the midpoint of the rising part at 16.5 K. The presence of a prominent *λ*-like anomaly around 16.5 K supports the existence of long-range magnetic ordering with a well-defined magnetic structure. This feature in *C*(*T*) is gradually smeared with the application of a magnetic field. It is seen that there is a weak, but distinct shift of the *λ*-peak towards higher temperature range with the increase in magnetic field, as evident from a comparison of the curves for 10 and 30 kOe with that of the zero field. This establishes that the long-range magnetic structure, setting in at 16.5 K, is of a ferromagnetic type. There is also a shoulder around 10 K in the zero-field curve, which is more transparent in the plot of $C/T$ (Figure 5b). This gets shifted towards lower temperature range with increasing $H$, which is clearly evident from the sudden change in the vicinity of 5 K in the curves for 50 and 100 kOe. Therefore, there is a support for the onset of an additional weak antiferromagnetic component around 10 K (in zero-field). It is not clear to us at present whether both ferromagnetic (indicated by magnetic hysteresis loop behavior) and antiferromagnetic phases coexist at a microscopic level or the ferromagnetism is of a weakly canted type with crystalline anisotropy playing a role. It is noted that $C(T)$ in zero field exhibits a power-law behavior ($T^{1.5}$) well below 10 K (figure 5(a), inset), different from $T^3$ dependence expected for a three-dimensional antiferromagnets, which indicate complexities of antiferromagnetism and/or spin-glasses [49]. Above $T_C$, there is no evidence for long-range ordering. It appears that (zero-field) $C/T$ curves tend to show a broad peak centered around 40 K, the origin of which is usually attributed to short-range magnetic clusters and/or crystal-field effects, as indicated by magnetization data. This can be inferred even from the $C(T)$ data after subtracting the lattice part, using the $C(T)$ data of La analogue [39] following the procedure suggested in Ref. 50; that is, this broad peak is distinctly visible in the magnetic contribution ($C_m$) to heat-capacity (Figure 5(c)). We have employed the heat-capacity data to derive isothermal entropy change, $\Delta S$ [defined as $S(H) - S(0)$], from the area under the curves of $C/T$ and the plots thus obtained are shown in Figure 5(d). A well-defined peak at the onset of long-range magnetic order is visible at 16.5 K; it is important to note that the sign of $\Delta S$ is negative at all fields, supporting [51] the conclusion that ferromagnetism sets in at 16.5 K. The fact that the tail of these entropy curves (at the right side of the peak) extends over a wide temperature range above 16.5 K at higher fields, suggests a gradual formation of ferromagnetic clusters before entering long range magnetically ordered state.

*3.3 ac magnetic susceptibility and isothermal remnant magnetization*

Since both antiferromagnetic and ferromagnetic correlations are present, we have explored the possibility of spin-glass dynamics by measuring *ac* $\chi$ as a function of $T$ with different frequencies. Such additional experiments are usually desirable, considering that the separation of low-field ZFC and FC $\chi$ curves can also arise in anisotropic magnetic materials, depending on relative magnitudes of the measuring field and the coercive field, as discussed in Ref. 52. The results are shown in figure 6.

In the zero-field measurements, at $T_C$, there is a peak at 16.5 K in the real part ($\chi'$), but the peak temperature undergoes a very weak change with increasing $\upsilon$, as indicated by an upward shift of the curve below the peak. The intensity of this peak decreases with increasing $\upsilon$. In addition, a peak appears in the imaginary part ($\chi''$) with the peak temperature (15.5 K) marginally lower than that at 16.5 K for $\chi'$. These peaks are completely destroyed when measured in a field of 5 kOe. All these features are characteristic of spin-glasses [42]. A notable observation is that relatively weaker, but $\upsilon$-dependent, broad peaks are observed even above 16.5 K near 20 and 25 K, suggesting that the clusters are formed before long range magnetic order, behaving like spin-glasses. Magnetic impurity contributions, if present, are not responsible to these additional anomalies above 16.5 K, considering that the features are sufficiently intense compared to the main peak and these extra phases cannot escape detection by x-ray diffraction and scanning electron microscope. Viewing together with the conclusions from the heat-capacity data presented above, one can argue that ferromagnetism arises from clusters and the inter-cluster dynamics is of spin-glass type. The aspect of cluster magnetism will be elaborated in subsequent discussions.

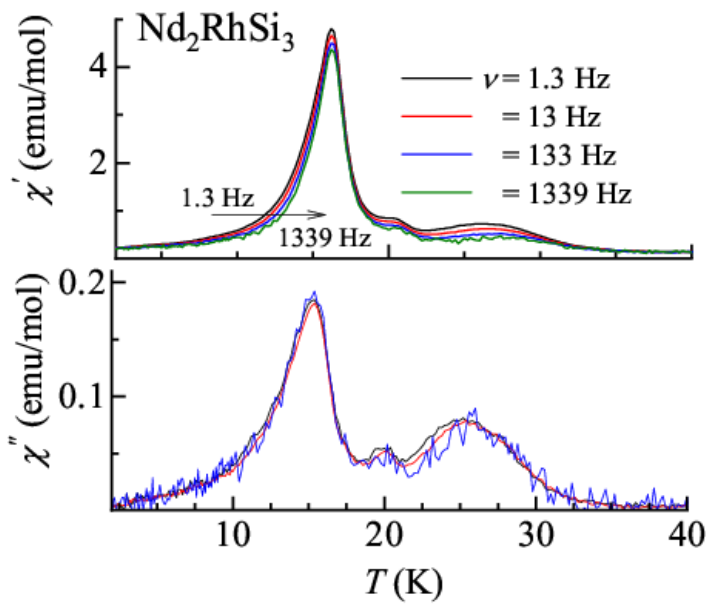

**Figure 6**. Real (top) and imaginary (bottom) parts of ac magnetic susceptibility for $Nd_2RhSi_3$ measured with 1.3, 13, 133, and 1339 Hz in the absence of external magnetic field. A horizontal arrow is drawn to show the direction in which the curve moves with increasing frequency. The $\chi''$ curves are found to be noisy in the higher frequency range and hence not shown for 1339 Hz. The peaks are suppressed when measured in an external field of 5 kOe (and hence not shown).

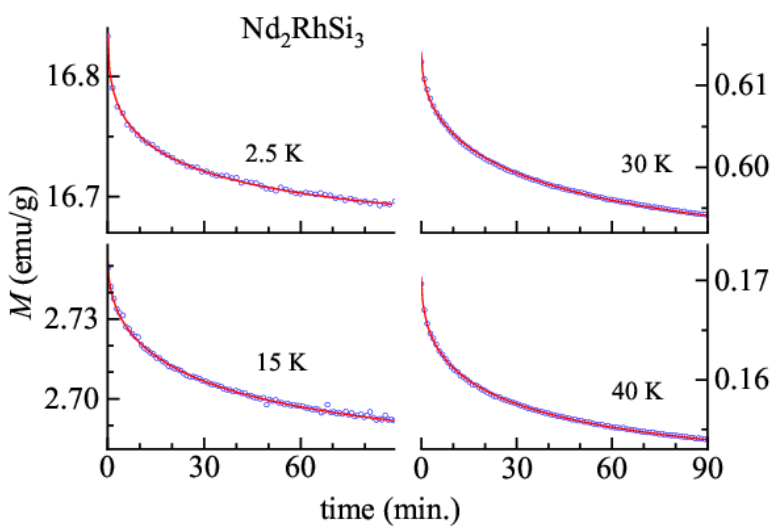

**Figure 7**. Isothermal remnant magnetization ($M_{IRM}$) curves as a function of time ($t$), obtained as described in the text, at 2.5, 15, 30 and 40 K. Continuous lines through the data points are obtained by a fit to a stretched exponential form.

To offer further support to the existence of spin-glass component, we show $M_{IRM}$ measured as a function of t in figure 7 at selected temperatures, not only below $T_C$, but also above. The value of $M_{IRM}$ is quite significant at 2.5 K (well below $T_C$) as soon as the field is switched off, undergoing slow decay with $t$, as expected for spin-glasses. The magnitude at $t = 0$ keeps decreasing with increasing $T$, as though spin-glass component becomes gradually weaker. Interestingly, a weak decay can be seen even at 40 K, consistent with the conclusions from ac $\chi$ data above. The curves could be fitted to a stretched exponential form [53], $M_{IRM}(t) = M_{IRM}(0) + A \exp(-t/\tau)^{1-n}$, where the constant A, the time constant $\tau$ and the exponent n are known to be related to the relaxation rate of the cluster spin-glass phase at respective temperatures. The fit revealed that the magnitude of $\tau$ falls in the range 1400 to 2600 s, typical of cluster spin-glasses [43-48], but it is found to vary non-monotonically with temperature attributable to complex nature of the magnetism. Corresponding non-monotonic variation of the exponent (falling in the range 0.4 -0.5) also could be seen.

*3.4 Electrical resistivity and magnetoresistance*

In the top panel of the figure 8, $\rho(T)$ plots in various fields are shown below 55 K only, as the curves in the range 40-300 K are featureless, with the slope in this temperature interval being positive, typical of metals. A sudden drop in $\rho$ (in the zero-field curve) is apparent at 16.5 K due to the loss of spin-disorder contribution, which strongly supports the conclusion from the $C(T)$ data that long range magnetic order with a well-defined magnetic structure sets in at this temperature. This drop persists even at higher fields, say in 10 and 30 kOe; a slope change can be distinctly seen even in 50 kOe. A linear extrapolation of the data points around this temperature revealed that there is an indeed a weak upward shift of the ordering temperature, characteristic of ferromagnetic ordering. This is also obvious from the derivate curve in figure 8 (bottom panel), and the temperature where a sudden change of slope (above 15 K) occurs corresponds to $T_C$. Spin-glass freezing alone would have smeared the feature around 16.5 K. There is no other worthwhile drop/peak above 16.5 K, consistent with the proposal that the *ac* $\chi$ features mentioned above at slightly higher temperatures are due to glassy magnetism. At lower temperatures (<6 K), $\rho$ exhibits quadratic *T*-dependence (Figure 8, inset), as though transport is dominated by ferromagnetism [54].

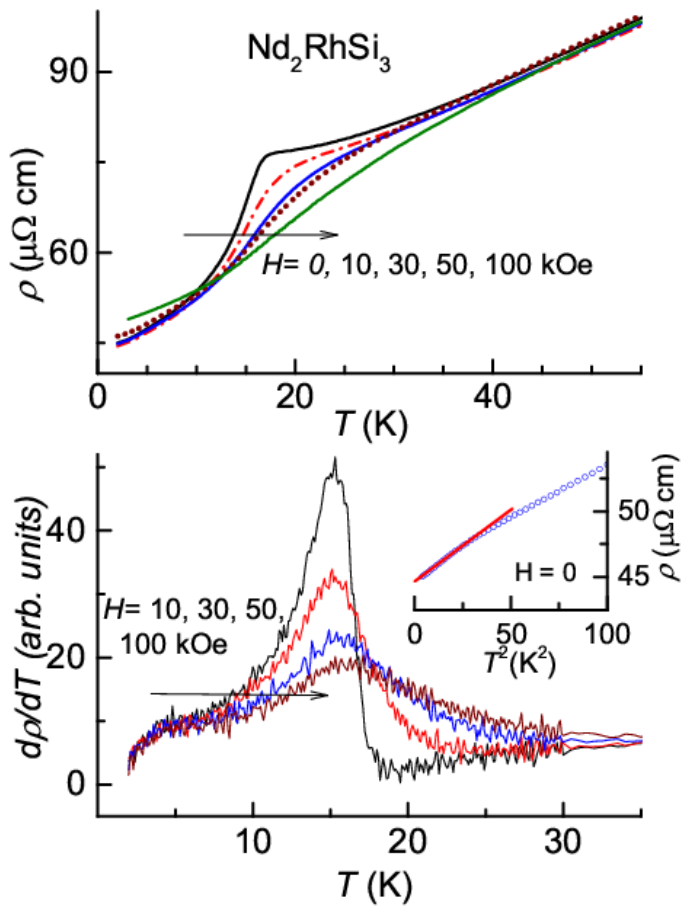

**Figure 8**. (Top) Electrical resistivity of $Nd_2RhSi_3$ as a function of temperature in 0, 10, 30, 50 and 100 Oe. A horizontal arrow is drawn to show the direction in which the curve (as seen at the magnetic transition) moves with increasing magnetic field. (Bottom) Derivative curves are shown to infer magnetic ordering temperature from the temperature where an increase in the slope occurs to infer how the magnetic ordering temperature changes with the magnetic field (also see the horizontal arrow). Inset shows the plot of (zero-field) resistivity as a function of temperature square and a line is drawn through the linear region at very low temperatures

Above ~6 K, another linear region can be seen and this deviation is attributed to the presence of an antiferromagnetic component. Thus these transport results support complex nature of magnetism at low temperatures. It can be inferred from the curves shown in figure 8 that the magnitude of MR keeps increasing with decreasing temperature below 50 K, similar to the magnetic precursor effects [55] seen in many other rare-earth systems. We have also probed MR behavior as a function of H at selected temperatures (figure 9) to get a reliable idea of sign reversals. At 2, 5, and 10 K, the sign of MR [defined as $\{\rho(H)-\rho(0)\}/\rho(0)$] is negative for initial applications of $H$ in the forward cycle, consistent with ferromagnetism or spin-glass; as the field is increased, there is a sign crossover with almost a quadratic field-dependence, which is attributed to the dominance of classical contribution from conduction electrons.

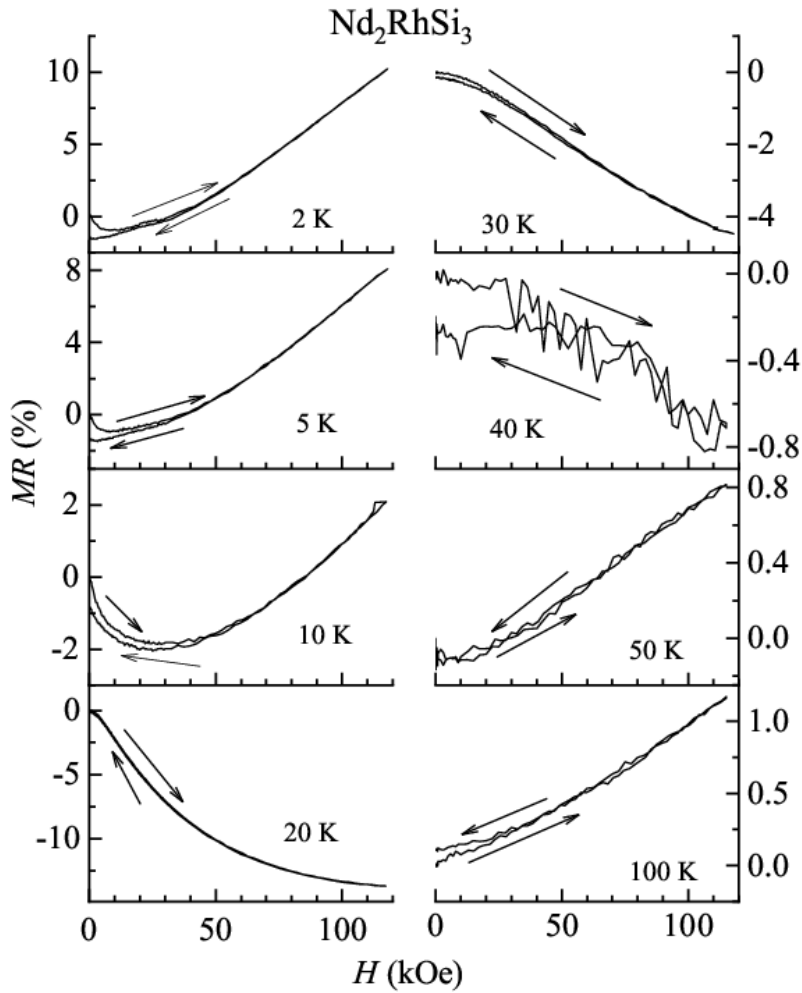

**Figure 9**. Magnetoresistance [MR= $\{\rho(H)-\rho(0)\}/\rho(0)$] as a function of magnetic field for Nd$_2$RhSi$_3$ at 2, 5, 10, 20, 30, 40, 50 and 100 K. The arrows serve as guides to show the way the magnetic field is varied.

When the field is reversed, a weak hysteretic behavior is observed as the field is approached towards zero, consistent with $M(H)$ behavior. Just above $T_C$, say, at 20 and 30 K, ferromagnetic cluster component overrides the classical contribution, which anyway should weaken with increasing $T$. Once the cluster glass region is almost crossed with increasing $T$, the classical contribution should dominate, leading to positive sign of MR, as observed experimentally at 50 and 100 K. Near 40 K, both the positive and negative contributions seem to compensate, leading to a negligible small value of MR (<<1%). (Given the small magnitudes of MR, we would claim that the hysteresis is absent at such high temperatures). In short, this compound exhibits an interesting interplay between the positive and negative contributions to MR. This behavior is somewhat similar to that observed for $Nd_2PdSi_3$.

## 4. Discussions

We now summarize main similarities and the differences with $Nd_2PdSi_3$ in the measured properties. Both the compounds show the features (e.g., in FC $\chi$, and the values of relaxation times) attributable to the onset of long-range ferromagnetic order with a cluster spin-glass-type dynamics below 16 K. Non-saturation of high-field magnetization in the magnetically ordered and the presence of an additional magnetic anomaly well below 16 K, revealing complexities in magnetism, are also similar. But the virgin curve behavior (lying inside the hysteresis loop) in the magnetic hysteretic curve of 1.8 K, unlike for the Pd case, establishes dominating ferromagnetic component for the Rh case. Following are major dissimilarities above $T_C$: (i) Persistence of open magnetic hysteretic loops (though weak) and, most notably, (ii) the separation of ZFC-FC $dc$ $\chi(T)$ curves and frequency dependent $ac$ $\chi$ features over a temperature range above $T_C$. These differences reveal gradual formation of magnetic clusters with spin-glass dynamics with a lowering of temperature towards $T_C$ for the Rh case only. Though it is irrelevant whether the clusters are of antiferromagnetic type or ferromagnetic type, the persistence of open hysteresis loop (above $T_C$) and the $T$-dependence of magnetic entropy data, support ferromagnetic nature.

We made an intriguing observation by comparing the long-range magnetic ordering temperatures of rare-earths in Pd and Rh ternary series. The Néel temperatures for Rh(Pd) family are, for instance: For Ce [6, 8, 30, 34, 35], 3 K (6 K); for Gd [3, 5, 39], 13.5 K (21 K); for Tb [4, 7, 12, 39], 13.5 K (23.6 K) ; for Dy [4, 39], 6.5 K (8 K); for Ho [10, 40] 5 K (8 K); and for Er [13, 40], 5 K (7 K). Thus, the long-range magnetic ordering temperatures of the $R$ ions (other than Nd) are by and large higher for the Pd series, compared to respective values for the Rh series. This implies that $R$ $4f$ interaction with Pd/Rh 4d dominates exchange interaction in these systems. However, in the case of the Nd compounds, this value is essentially the same (~16 K) in both the series, despite that Pd is replaced by Rh. Note that ferromagnetic ordering takes place in both the compounds, despite that, for all other members including Ce, the onset of magnetic ordering is of an antiferromagnetic-type. It is expected that $4f$ hybridization is stronger for Ce cases than that for corresponding Nd cases, and if $4f$ hybridization with Pd/Rh $4d$ orbital is responsible for anomalous magnetism, the magnetic behavior of Ce in both the series should be similar to that of respective Nd compound, in contrast to the observation. Therefore, in the Nd compounds, from the above comparison with the magnetic behavior of Ce cases, one sees an agreement with the conclusion made in Ref. 18 that $4f$ hybridization with $3p$ orbitals of Si is the decisive factor to make the magnetism of Nd compounds different from other members of these two ternary families. In the case of $R_2PtSi_3$ as well, limited information available in the literature within this series [56, 57] brings out that the $T_C$ of Nd compound (17.5 K) is very close to that of the above two Nd compounds, whereas, for the Gd case, the corresponding value is significantly enhanced (to 24 K), supporting the above inference. However, the differences - ferromagnetic-like $M(H)$ loop for the Rh case at very low temperatures in contrast to dominating antiferromagnetic-like loop for the Pd case, and clear evidence for cluster-glass anomalies in the $\chi$ data above $T_C$ for the former - cannot be ignored. We infer that the hybridization with d band of transition metals plays a role (in terms of geometrically frustrated magnetism, see also below). The fact is that the Nd compounds of $AlB_2$-derived structure have not been subjected to such exhaustive comparative studies carefully (in some cases due to the absence of all members in a given $R$ series) to explore intricacies of Nd 4*f* hybridization. Magnetic investigations on a few members of $R_2CuSi_3$ [Refs. 58-60], $R_2AgIn_3$ [Refs. 61-63], $R_2CuIn_3$ [Ref. 64], etc. have been sporadically reported, revealing interesting magnetic characteristics of Nd compounds, including significant enhancement of magnetic ordering temperature with respect to de Gennes scaling. It is of interest to note that the real space high resolution transmission electron spectroscopic studies [59] on Ce and Nd analogues in the family $R_2CuSi_3$ revealed relatively large sizes of clusters for the latter compared to the former. It is therefore of interest to carry out similar studies on the title compound as well as small angle neutron diffraction studies as a function of temperature down to 2 K to get a better idea of cluster magnetism. [A large number of families (about 46 structure types [65]) can be derived from $AlB_2$-type structure, and therefore, exhaustive investigations of all these compounds would be rewarding in the context of the conclusions on this Nd compound].

Several magnetic phases forming such clusters competing as discussed above with a gradual lowering of temperature towards magnetic ordering could be related to geometrical frustration in, say, a triangular network of magnetic ions, as emphasized for $Tb_3Ru_4Al_{12}$ [43], consistent with different theoretical approaches by Wang et al. [21], Jaubert et al. [66], Schmidt et al. [67], and Silveira et al. [68]. Such a magnetic state appearing before long-range magnetic order has been called by different terminologies, e.g., "correlated paramagnetic state" [69], "classical spin liquid state" [21, 68] and, more broadly "magnetic precursor effects" [55].

As stated earlier, a neutron diffraction studies [40] at 4.2 K suggested spiral ferromagnetic ordering. Thus, these results may be interpreted that $Nd_2RhSi_3$ serves as a rare demonstration for spin-glass dynamics of spiral ferromagnetic structure. However, considering the outcome of recent neutron diffraction studies [19] on isostructural $Nd_2PdSi_3$, the physics of the present compound may not be that simplistic and it is tempting to look at an alternate microscopic interpretation for spin glass dynamics. In this connection, it is important to note that the triangular arrangement in this family is quite revealing. As stated at the Introduction, the lattice is quite strained with multiple bond distances for a given pair of atoms, closely related to the doubling of the unit cell with respect to primitive $AlB_2$ hexagonal structure. It is clear from the figure 1b that the triangular network of *R* is made up of triangles of slightly different dimensions and there are non-equilateral triangles as well. This means that the rare-earth layer is considerably strained, leading to bond disorder. This is a kind of geometry-induced bond disorder or random strains, which naturally results in randomness in exchange interactions; this should cause microscopic magnetic phase separation (even in a stoichiometric compound) into 'partially disordered magnetic' and long range ordered components with the former behaving like a cluster spin-glass. Such a scenario is possible in $Er_2RhSi_3$ as well, in which case spin-glass component has been shown recently [39] to coexist with antiferromagnetic order. Bond-disorder induced magnetic frustration has been theoretically predicted [70, 71] to lead to spin-glass features. Therefore, in our opinion, this is a likely microscopic scenario to understand spin-glass dynamics.

## 5. Conclusions

In summary, we have investigated the detailed electronic properties of $Nd_2RhSi_3$ employing bulk experimental methods. The results establish that this compound exhibits long range ordering of a ferromagnetic type (at 16. 5 K) with a concomitant cluster spin-glass features, and also with evidences for the gradual onset of such clusters as the temperature is lowered towards $T_C$. It is of interest to investigate the role of possible Griffiths phase formation in these cluster magnetic anomalies. The magnetism is complex at temperatures lower than ~10 K with the development of an antiferromagnetic component. Crystallography-favored bond disorder, coupled with local variations in Nd 4*f* covalency, should be a key for such a frustrated magnetism. Detailed neutron diffraction studies as a function of temperature, as in Nd$_2$PdSi$_3$ [19] extending the work of Ref. 72, would be required to understand magnetism of this compound better.

Studies by small angle neutron diffraction, inelastic neutron scattering, muon spin rotation techniques and electronic structure as a function of temperature down to 2 K, as well as investigations on single crystals, would also be rewarding. In any case, the observation of dominant ferromagnetic behavior, different from that in other isostructural members of this Rh family, which have been known to undergo antiferromagnetic ordering and the gradual development of magnetic clusters with lowering temperatures uncharacteristic of other members of such ternary families including Nd$_2$PdSi$_3$, make this Nd compound an exotic magnetic material. This work firmly establishes that the direction of careful investigations of Nd based compounds to look for anomalous 4*f*-hybridization related magnetic anomalies would be a rewarding avenue. The discovery of such compounds would contribute [73] to a better development of the field of magnetic skyrmions.

## Acknowledgements

Authors acknowledge financial support from the Department of Atomic Energy (DAE), Govt. of India (Project Identification no. RTI4003, DAE OM no. 1303/2/2019/R&D-II/DAE/2079 dated 11.02.2020). One of the authors (E.V.S) would like to thank Department of Atomic Energy, Government of India, for awarding Raja Ramanna Fellowship to carry out this work.